\documentclass[aps,prb,twocolumn,groupedaddress,showpacs]{revtex4}
\usepackage{epsfig}
\usepackage{color}
\begin{document}

\title{Quantum Monte Carlo Study of a Dynamic Hubbard Model}
\author{K. Bouadim$^1$, M. Enjalran$^2$, F. H\'ebert$^1$,
G.G. Batrouni$^1$, and R.T. Scalettar$^3$}
\affiliation{$^1$INLN, Universit\'e de Nice-Sophia Antipolis, CNRS;
1361 route des Lucioles, 06560 Valbonne, France}
\affiliation{$^2$Department of Physics,
Southern Connecticut State University,
New Haven, CT 06515-1355}
\affiliation{$^3$Physics Department, University of California, 
Davis, California 95616, USA}

\begin{abstract}
The `dynamic' Hubbard Hamiltonian describes interacting fermions on a
lattice whose on-site repulsion is modulated by a coupling to a
fluctuating bosonic field.  We investigate one such model, introduced
by Hirsch, using the determinant Quantum Monte Carlo method.  Our key
result is that the extended $s$-wave pairing vertex, repulsive in the
usual static Hubbard model, becomes attractive as the coupling to the
fluctuating Bose field increases. The sign problem prevents us from
exploring a low enough temperature to see if a superconducting
transition occurs.  We also observe a stabilization of
antiferromagnetic correlations and the Mott gap near half-filling, and
a near linear behavior of the energy as a function of particle density
which indicates a tendency toward phase separation.
\end{abstract}

\pacs{
71.10.Fd, % Lattice fermion models (Hubbard model, etc.)
71.30.+h, % Metal-insulator transitions and other electronic transitions
02.70.Uu  % Applications of Monte Carlo methods
}
\maketitle

\section{Introduction}

The fermion Hubbard Hamiltonian \cite{fazekas99}, originally proposed
to describe the physics of transition metal monoxides FeO, MnO, and
CoO, has been widely used as a model of cuprate superconductors, whose
undoped parent compounds, like La$_{2}$CuO$_4$, are also
antiferromagnetic and insulating.  Indeed, early Quantum Monte Carlo
(QMC) simulations of the Hubbard Hamiltonian suggested that d-wave
pairing was the dominant superconducting
instability\cite{white89,white89b}, a symmetry which was subsequently
observed in the cuprates\cite{dwaveexpt}.  However, the sign problem
precluded any definitive statement about a phase transition to a
d-wave superconducting phase\cite{white89b,loh90}.  Over the last
several years, QMC studies within dynamical mean field theory and its
cluster generalization \cite{maier00,maier06} are presenting a more
compelling case for this transition.  The existence of charge
inhomogeneities in Hartree-Fock \cite{zaanenxx} and density matrix
renormalization group treatments\cite{whitexx}, along with the
experimental observation of such patterns \cite{chargeinhomoexpt},
offer further indications that significant aspects of the qualitative
physics of the cuprates might be contained in the Hubbard Hamiltonian.

Nevertheless, there are a number of features of high temperature
superconductors which do not completely fit within the framework of
the single band Hubbard Hamiltonian.  For example, the cuprate gap is
set by the charge transfer energy separating the copper $d$ and oxygen
$p$ orbitals \cite{emery87,zhang88} as opposed to a Mott gap between
copper $d$ states split by the on-site repulsion.  Considerable
evidence for the possible important role of phonon modes in aspects of
the physics is available\cite{phonons}.

Hirsch has emphasized the asymmetry in transition temperatures, and
other properties, between the electron and hole doped cuprates as a
reason to consider more general models, since the particle-hole
symmetry of the single band Hubbard Hamiltonian requires that its
behavior be rigorously identical for fillings $\rho=1-x$ and
$\rho=1+x$.  Partially motivated by this asymmetry, he introduced
\cite{hirsch01,hirsch02,hirsch02b,hirsch03,marsiglio03} the dynamic
Hubbard Hamiltonian,
\begin{eqnarray}
H&=& -t \sum_{\langle \bf{i},\bf{j} \rangle} \big(
c^\dagger_{\bf{j}\sigma} c_{\bf{i}\sigma} + c^\dagger_{\bf{i}\sigma}
c_{\bf{j}\sigma} \big) - \mu \sum_{\bf{i}} (n_{\bf{i}\uparrow} +
n_{\bf{i}\downarrow}) \nonumber \\ &+& \sum_{\bf{i}} \Big[\omega_0
\sigma^x_{\bf{i}} + g \omega_0\sigma^z_{\bf{i}} + ( \, U-2 g \omega_0
\sigma^z_{\bf{i}} \,) \, n_{\bf{i}\uparrow}n_{\bf{i}\downarrow} \Big]
.
\label{Hamilton}
\end{eqnarray}
Here the first term, involving the fermion creation (destruction)
operators $c^{\dagger}_{\bf{j} \sigma} (c_{\bf{j} \sigma})$ at site
$\bf{j}$ with spin $\sigma$, is the tight binding kinetic energy describing
the hopping of electrons between near neighbor sites.  We consider
here a two-dimensional square lattice and chose $t=1$ to set our scale
of energy.  The on-site interaction energy differs from the usual
static Hubbard Hamiltonian in that its value $U$ is modulated by a
dynamic field $\sigma^z_{\bf{i}}$ which can take the values
$\sigma^z_{\bf{i}} = \pm 1$.  As a consequence the on-site repulsion
$U$ has bimodal values $U_{\rm min}=U-2 g \omega_0$ and $U_{\rm
max}=U+2 g \omega_0$.  This dynamic field itself has non-trivial
quantum fluctuations controlled by the relative values of the
longitudinal and transverse frequencies $g \omega_0$ and $\omega_0$.
Here $\sigma^x_{\bf{i}}$ and $\sigma^z_{\bf{i}}$ are Pauli
matrices\cite{note1}.  The variation in $U$, Hirsch argued, has its
physical origin in the relaxation which occurs with multiple
occupation of an atomic orbital.

Hirsch and collaborators have studied the physics of
Eq.~\ref{Hamilton} with a variety of methods, including a Lang-Firsov
transformation (LFT)\cite{hirsch01}, exact diagonalization (ED) of
small clusters\cite{hirsch02b,hirsch03}, and world-line Quantum Monte
Carlo (WLQMC) \cite{hirsch82} in one dimension \cite{hirsch02}.
Within the LFT it is seen that the hopping of electrons is
renormalized by the overlap of the states of the dynamic variable on
neighboring lattice sites.  Superconductivity then arises because
isolated holes are essentially localized by a small overlap, whereas
holes that are on the same or neighboring sites can move around the
lattice.  Furthermore this effect is operative for holes in a nearly
filled system, but not electrons in a nearly empty lattice.  Thus
pairing is linked to the presence of holes, and the physics is
manifestly not particle-hole symmetric.  ED provided quantitative
values for the overlaps and confirmed the picture based on the LFT on
small clusters.

\begin{figure}
\centerline{\epsfig{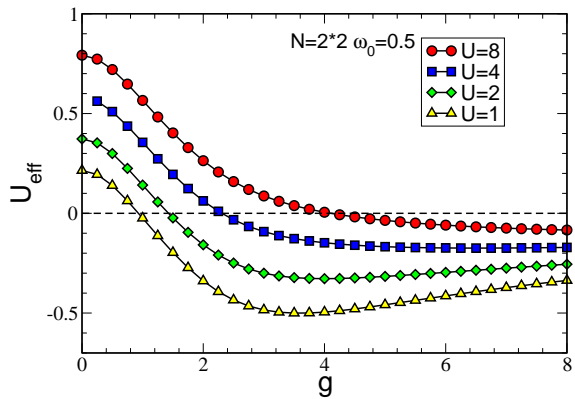}}
\caption{(Color online) Binding energy $U_{\rm eff}$ on a four site
cluster as a function of the coupling $g$ to the dynamically
fluctuating field $\sigma^z$.  The binding energy $U_{\rm eff}$ can go
negative at large $g$, suggesting the possibility of pairing.  }
\label{hirschcheck1}
\end{figure}

ED also allows for the evaluation of the `binding energy', $U_{\rm
eff}=2E_0(N+1) - E_0(N+2)-E_0(N)$.  Here $E_0(N)$ is the ground state
energy of a cluster with $N$ electrons.  A negative $U_{\rm eff}$
indicates that it is energetically favorable to put two particles
together on a single cluster rather than separate them on two
different clusters. On a sufficiently large lattice, two particles
would tend to be close spatially rather than widely separated.  In
Fig.~\ref{hirschcheck1} we show an evaluation of $U_{\rm eff}$ on a
2x2 lattice.  These numbers were obtained independently from, but
are identical to, those of
Ref.~\onlinecite{hirsch02}. As the coupling $g$ to the dynamic field
increases, $U_{\rm eff}$ is driven negative, indicating the
possibility of binding of particles and hence superconductivity.
WLQMC simulations in one dimension confirmed this real space pairing
by explicitly showing the preference of the world lines of holes to
propagate next to each other and a large gain in kinetic energy when
the hole-hole separation becomes small.  Significantly, these
simulations also showed that the kinetic energy disfavors proximity of
holes in the Holstein model, which also features the tendency of holes
to clump together by distorting a local phonon degree of freedom.
Thus pairing in the dynamic Hubbard model is distinguished from that
of more traditional electron-phonon models by being driven by the
kinetic energy as opposed to a potential energy.

In this paper we examine the properties of the dynamic Hubbard
Hamiltonian with determinant Quantum Monte Carlo
(DQMC)\cite{blankenbecler81}. This approach allows us to work in two
dimensions, as opposed to previous ($d=1$) WLQMC studies, and also to
examine lattices of an order of magnitude greater number of sites than
ED.  On the other hand, the ability of DQMC to reach low temperatures
is limited by the sign problem\cite{loh90}.  We find that the extended
$s$-wave pairing vertex, which is repulsive in the static Hubbard
model, is attractive in the dynamic model, that is,
extended $s$-wave superconducting correlations are
 enhanced by the dynamic fluctuations.  However, the pairing
susceptibilities are still only rather weakly increasing down to the
lowest temperatures accessible to us (temperature $T$ greater than
1/40 the electronic bandwidth).

We also find, near half-filling, that the antiferromagnetic
correlations can be enhanced relative to the static Hubbard
Hamiltonian, particularly for densities {\it above} $\rho=1$.  The
Mott gap can also be stabilized.  Interestingly, the total energy
appears to be close to linear in the particle density, as opposed to a
clear concave up curvature in the static Hubbard model (with either
repulsive or attractive interactions).

The organization of this paper is as follows: In the next section we
present our computational method, DQMC, as it applies to the dynamic
Hubbard model. We describe several minor adjustments to the DQMC
algorithm for the static Hubbard model that are needed in order to
study the dynamic model. Our observables are also defined.  In Section
III, we present the results from our Monte Carlo simulations.  The
topics of antiferromagnetism and the Mott transition, pair
susceptibilities and superconductivity, and the energy characteristics
of the dynamic Hubbard model are discussed. The paper closes with
conclusions in Section IV.

\section{Computational Methods}

Although he did not undertake such studies, Hirsch pointed out
\cite{hirsch02} that the dynamic Hubbard model could be simulated with
a relatively minor modification of the DQMC
method\cite{blankenbecler81}.  In DQMC, an auxiliary
`Hubbard-Stratonovich' (HS) field is introduced to decouple the
on-site Hubbard repulsion.  The trace over the resulting quadratic
form of fermion operators is performed analytically, leaving an
expression for the partition function which is a sum over the HS
variables whose weight is given by the product of two determinants,
one for spin up and one for spin down, that are produced by evaluating
the trace.

In DQMC for the usual Hubbard Hamiltonian, the HS field couples to the
difference between the up and down spin
electron densities, with a coupling constant which is independent of
spatial site and imaginary time.  In a simulation of the dynamic
Hubbard model, the coupling of the HS field depends on the dynamic
field $\sigma^z_{\bf{i}}(\tau)$.  The imaginary time dependence arises
from the transverse term $\sigma^x_{\bf{i}}$ in the Hamiltonian.  When
the path integral for the partition function is constructed
$\sigma^x_{\bf{i}}$ induces flips between the two values
$\sigma^z_{\bf{i}}= \pm 1$, so this quantity becomes dependent on
$\tau$.  As a consequence, minor modifications are required to the
standard expressions \cite{blankenbecler81} for the ratio of
determinants before and after the Monte Carlo move of a HS variable,
and for the re-evaluation of the Green's function.

The dynamic field variables must also be updated, and again, only
minor modifications of the formulae for the determinant ratio and
Green's function update are required.  A final difference is that
there is a contribution to the weight coming from the $\sigma^x$ and
$\sigma^z$ terms in the Hamiltonian.  The former try to align the
dynamic variables in the imaginary time direction, while the latter
favor positive values of the dynamic field.  Such pieces of the
action, which enter the weight of the configuration along with the
fermion determinants, are similar to those arising in simulations of
the Holstein Hamiltonian\cite{niyaz93}.

\begin{figure}
\centerline{\epsfig{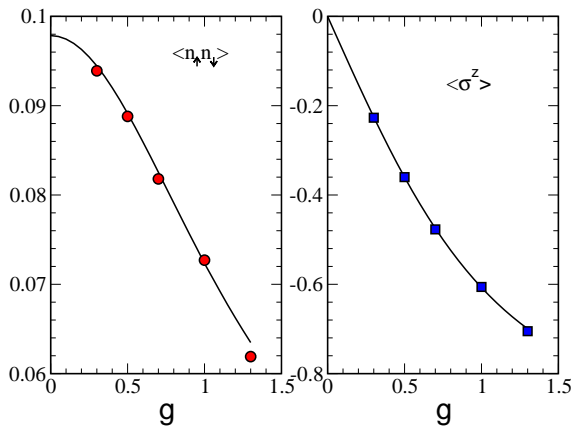}}
\caption{(Color online) Double occupancy (left) and expectation value
of dynamic field (right) as functions of the coupling $g$.  The solid
line is the result of exact diagonalization and the symbols of the
determinant QMC simulations.  The cluster size is 2x2 (the same as for
the binding energy calculation of Fig.~\ref{hirschcheck1} and
Ref.~\onlinecite{hirsch02}).  Parameters are $t=1, \, U=4, \,
\beta=1.3, \, \mu=2$ and $\omega_0 =1.5$.  }
\label{sdcheck}
\end{figure}

We verified our DQMC code by comparing to exact diagonalization
results on a 2x2 spatial lattice
(Figs.~\ref{hirschcheck1},\ref{sdcheck}), and also by checking
analytically soluble limits such as $t=0$.  The results of our
DQMC/diagonalization calculations on 2x2 lattices are completely
consistent with those of Hirsch.  For example, we have quantitatively
reproduced the binding energy plot, Fig.~1(top) of
Ref.~\onlinecite{hirsch02} and our Fig.~\ref{hirschcheck1}.  As a
further check, we compared DQMC results for the double occupancy,
$\langle n_\uparrow n_\downarrow \rangle$, and the expectation value
of the dynamic field, $\langle \sigma^z \rangle$, to results from ED.
See Fig.~\ref{sdcheck}.

We did not observe any major difference in the characteristics of the
DQMC algorithm in simulating the dynamic Hubbard model:
Autocorrelation times remain short, as is typically the case with
DQMC, and there was no major change in the numerical stability
\cite{white89b,sugiyama86,fye88,sorella89}. The key issue in DQMC is
the `sign problem' which we will discuss in the following sections.

DQMC allows us to measure any observable which can be expressed as an
expectation value of products of creation and destruction operators.
Our measurements include the energy $\langle H \rangle$ (not including
the chemical potential term), particle density $\rho = \langle n
\rangle$, and Green's function $G_{\bf{ij}}(\tau)= \langle
c_{\bf{i}}(\tau) c_{\bf{j}}^\dagger(0) \rangle$, as well as the
average of the dynamic field $\langle \sigma^z_{\bf{i}} \rangle$.  The
dependence of the density on the chemical potential $\mu$ and the
Green's function, when analytically continued to the spectral function,
allows us to examine, among other things, the Mott
metal-insulator transition.

In addition to these single particle properties we also examine
magnetic correlations, and specifically, the magnetic structure
factor,
\begin{eqnarray}
S({\bf k}) = \sum_{{\bf l}}e^{i{\bf k} \cdot {\bf l}} \,\, \langle
\,\, (n_{{\bf j+l} \uparrow } - n_{{\bf j+l} \downarrow } ) (n_{{\bf
j} \uparrow} - n_{{\bf j} \downarrow} ) \,\, \rangle \,\, .
\label{SDW}
\end{eqnarray}
Our focus will be on the antiferromagnetic response, $S({\bf
k}=(\pi,\pi))$.

We look at superconductivity by computing the correlated pair field
susceptibility, $P_{\alpha}$, in different symmetry channels,
\begin{eqnarray}
P_{\alpha} &=& \int_0^\beta d \tau \langle \Delta_{\alpha}(\tau)
\Delta_{\alpha}^\dagger(0) \rangle \nonumber \\
\Delta^\dagger_{\alpha} &=& {1 \over N} \sum_{\bf k} f_\alpha ({\bf
k}) c^\dagger_{{\bf k}\uparrow} c^\dagger_{-{\bf k}\downarrow}
\nonumber \\ f_s({\bf k}) &=& 1 \nonumber \\ f_{s^*}({\bf k}) &=& {\rm
cos} (k_x) + {\rm cos} (k_y) \nonumber \\ f_d({\bf k}) &=& {\rm cos}
(k_x) - {\rm cos} (k_y) \,\,.
\label{Pmom}
\end{eqnarray}
These quantities can also be expressed in real space,
\begin{eqnarray}
\Delta^\dagger_s &=& {1 \over N} \sum_{\bf i} c^\dagger_{{\bf i}\uparrow}
c^\dagger_{{\bf i}\downarrow}
\nonumber \\
\Delta^\dagger_{s^*} &=& {1 \over N} \sum_{\bf i} c^\dagger_{{\bf i}\uparrow}
\Big( c^\dagger_{{\bf i}+\hat x\downarrow}
+c^\dagger_{{\bf i}+\hat y\downarrow}
+c^\dagger_{{\bf i}-\hat x\downarrow}
+c^\dagger_{{\bf i}-\hat y\downarrow} \Big)
\nonumber \\
\Delta^\dagger_{d} &=& {1 \over N} \sum_{\bf i} c^\dagger_{{\bf i}\uparrow}
\Big( c^\dagger_{{\bf i}+\hat x\downarrow}
-c^\dagger_{{\bf i}+\hat y\downarrow}
+c^\dagger_{{\bf i}-\hat x\downarrow}
-c^\dagger_{{\bf i}-\hat y\downarrow} \Big)
\,\,.
\label{Preal}
\end{eqnarray}

The correlated susceptibility $P_\alpha$ takes the expectation value
of the product of the four fermion operators entering Eq.~\ref{Pmom}.
We also define the uncorrelated pair field susceptibility
$\overline{P}_\alpha$ which instead computes the expectation values of
pairs of operators {\it prior} to taking the product.  Thus, for
example, in the $s$-wave channel,
\begin{eqnarray}
P_{s} &=& 
{1 \over N^2} 
\sum_{{\bf i},{\bf j}}
\int_0^\beta d \tau 
\langle \,
c_{{\bf i}\downarrow}(\tau) \, c_{{\bf i}\uparrow}(\tau) 
\, c^\dagger_{{\bf j}\uparrow}(0) \, c^\dagger_{{\bf j}\downarrow}(0) 
\, \rangle 
\nonumber \\
\overline{P}_{s} &=& 
{1 \over N^2} 
\sum_{{\bf i},{\bf j}}
\int_0^\beta d \tau 
\langle \,
c_{{\bf i}\downarrow}(\tau) 
\, c^\dagger_{{\bf j}\downarrow}(0) 
\, \rangle  \,\,
\langle \,
c_{{\bf i}\uparrow}(\tau) 
\, c^\dagger_{{\bf j}\uparrow}(0) 
\, \rangle 
\,\,.
\label{Pbar}
\end{eqnarray}
$P_\alpha$ includes both the renormalization of the propagation of the
individual fermions as well as the interaction vertex between them,
whereas $\overline{P}_\alpha$ includes only the former effect.  Indeed
by evaluating both $P$ and $\overline{P}$ we are able to extract
\cite{getgamma} the interaction vertex $\Gamma$,
\begin{eqnarray}
\Gamma_\alpha = {1 \over P_\alpha}
- {1 \over \overline{P}_\alpha} \,\,.
\end{eqnarray}
If $\Gamma_\alpha < 0$, the associated pairing interaction is
attractive.  $\Gamma_\alpha \rightarrow -1$ signals a superconducting
instability.

\section{Results}

\subsection{Mott Transition and Antiferromagnetism}

\begin{figure}
\centerline{\epsfig{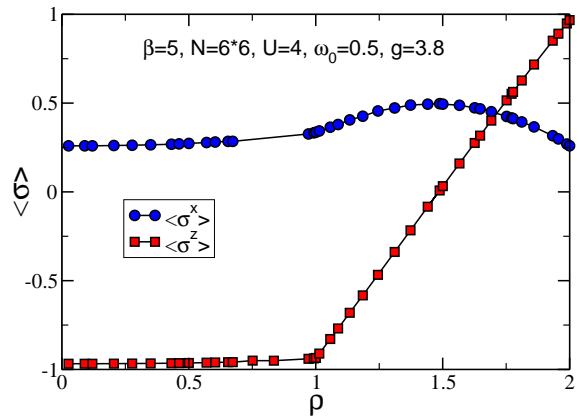}}
\caption{(Color online) $\langle\sigma^z\rangle$ and
$\langle\sigma^x\rangle$ as a function of $\rho$ for $\omega_0=0.5,
g=3.8$.  From $\rho=0$ to half-filling, the system minimizes its
energy by maximizing the interaction term $U-2g\omega_0 \sigma^z$ to
avoid double occupation, that is, $\langle \sigma^z \rangle \simeq
-1$.  In this Figure, and elsewhere in this paper, the lattice size is
6x6 unless otherwise stated.}
\label{Sigmaxzwz}
\end{figure}

It is useful to begin our study of the dynamic Hubbard model by
understanding the behavior of the dynamic field at different fillings
(Fig.~\ref{Sigmaxzwz}).  For fillings below one particle per site,
$\rho<1$, the dynamic field $\sigma^z_{\bf{i}} \simeq -1$ because 
of the coupling to the external field$g\omega_0$ hence the interaction 
$U+2g \omega_0 \simeq U_{\rm max}$ and the double occupancy is reduced.  
However, once double occupancy
is unavoidable ($\rho > 1)$, the interaction term strongly favors
$\sigma^z_{\bf{i}}=+1$.  Fig.~\ref{Sigmaxzwz} shows that this
evolution from negative to positive values is nearly linear once
$\rho>1$.  Meanwhile, the expectation value of $\sigma^x_{\bf{i}}$
measures the fluctuations of $\sigma^z_{\bf{i}}$ in imaginary time.
It is not surprising, then, that this quantity exhibits a maximum at
roughly the midpoint between the evolution from $\sigma^z_{\bf{i}}
=-1$ to $\sigma^z_{\bf{i}} =+1$, at $\rho \approx 1.5$.  In the
results of Fig.~\ref{Sigmaxzwz}, and throughout this paper unless
otherwise stated, the simulations were performed on 6x6 lattices.

\begin{figure}
\centerline{\epsfig{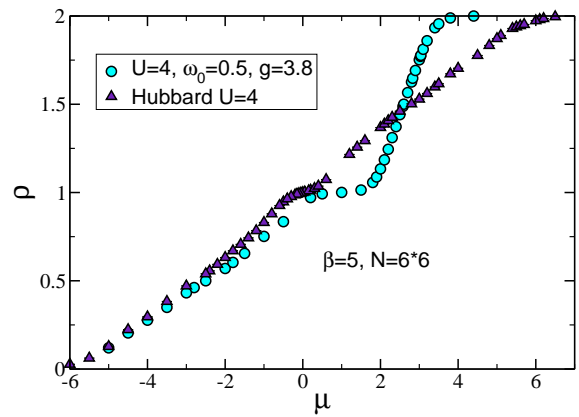}}
\caption{(Color online) Comparison of the evolution of the density
$\rho$ with chemical potential $\mu$ for the static and dynamic
Hubbard models.  The dynamic model has a significantly better
developed Mott insulating gap, as well as a pronounced particle-hole
asymmetry.  Here the inverse temperature $\beta=5$.}
\label{Murhobis}
\end{figure}

Next, we compare the Mott gap and magnetic correlations in the static
and dynamic Hubbard models.  In Fig.~\ref{Murhobis} we plot the
density $\rho$ as a function of chemical potential $\mu$.  A plateau
at $\rho=1$ indicates the formation of a Mott insulator.  The cost to
add a particle suddenly jumps by $U$ because additional particles are forced
to sit on sites which are already occupied.  At the
inverse temperature chosen, $\beta=5$, for the static Hubbard model,
the plateau is only beginning to develop.  However for the dynamic
model the plateau is much more robust.  This is expected since near
half-filling, as we have seen, the on-site repulsion mostly takes on
its maximum value $U_{\rm max} = 7.8$, for the parameters 
in Fig.~\ref{Murhobis}.  We have
chosen dynamic Hubbard parameters $g$ and $\omega_0$ which gets the
system as close as possible to the most attractive (negative) binding
energy $U_{\rm eff}$ while still having $U_{\rm min} > 0$.

\begin{figure}
\centerline{\epsfig{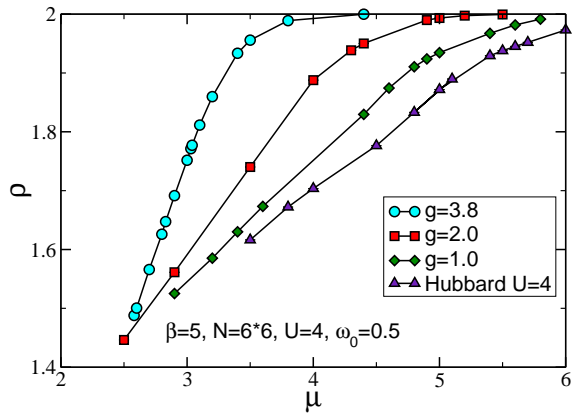}}
\caption{(Color online) The density $\rho$ as a function of chemical
potential $\mu$ at $U=4$ and $\beta=5$.  As the coupling $g$
increases, the cost to add particles to an already occupied site
decreases.  As a consequence, $\rho$ rises more steeply with $\mu$.}
\label{Murho}
\end{figure}

\begin{figure}
\centerline{\epsfig{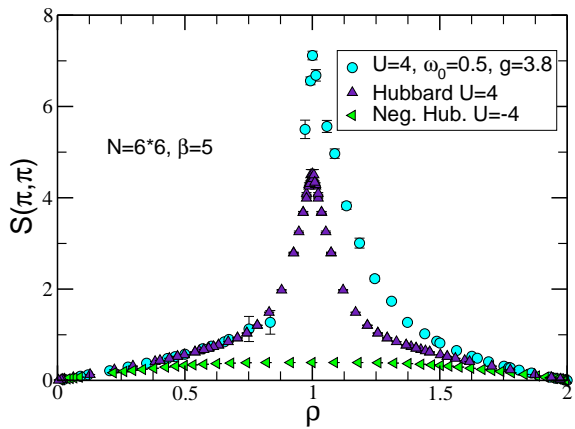}}
\caption{(Color online) The antiferromagnetic structure factor
$S(\pi,\pi)$ at inverse temperature $\beta=5$ as a function of density
$\rho$.  For both the static and dynamic repulsive Hubbard
Hamiltonians there is significant antiferromagnetic order near half
filling, with the magnetic correlations in the dynamic model somewhat
more robust.  There is no magnetic signal for the attractive model
which, instead, is known to show strong charge density wave and
$s$-wave superconducting correlations.  }
\label{AF}
\end{figure}

Figure~\ref{Murho} gives further insight into the behavior of the
density near full filling.  In the static model, the cost to add
particles to the system is set by the on-site $U$ (in the case that
$U$ exceeds the bandwidth $W=8t$).  However, in the dynamic model, as
full filling is approached, the double occupancy cost
is reduced to $U_{\rm min}$.  For the parameters chosen in
Fig.~\ref{Murho}, $U_{\rm min} = 0.2$ is close to zero.  Thus we
expect the filling of the lattice to be complete when the chemical
potential reaches the top of the band, $4t$, in good agreement with
the plot.

The static Hubbard model exhibits antiferromagnetic correlations at
half-filling on a bipartite lattice, since only electrons with anti-aligned
spins can hop between neighboring sites.  This leads to a lowering of
the energy by the exchange energy $J = 4t^2/U$ relative to sites with
parallel spin, for which hopping is forbidden.  Indeed, a finite size
scaling analysis of the structure factor has shown there is long range
order in the ground state\cite{hirsch89}.  Figure~\ref{AF} compares
the value of the antiferomagnetic structure factor $S(\pi,\pi)$ for
the static and dynamic models.  At half-filling, $S(\pi,\pi)$ for the
dynamic model attains a maximal value 50\% larger than that of the
static model.  There is a marked asymmetry in the magnetic response at
values greater and lower than $\rho=1$ in the dynamic model, with
$S(\pi,\pi)$ remaining high to values of $\rho$ ten percent larger
than half-filling.  We also show results for the negative $U$ Hubbard
model, which has no tendency for magnetic order at any filling.
(Instead, the attractive Hubbard model exhibits long range charge and
superconducting correlations at $\rho=1$).

\begin{figure}
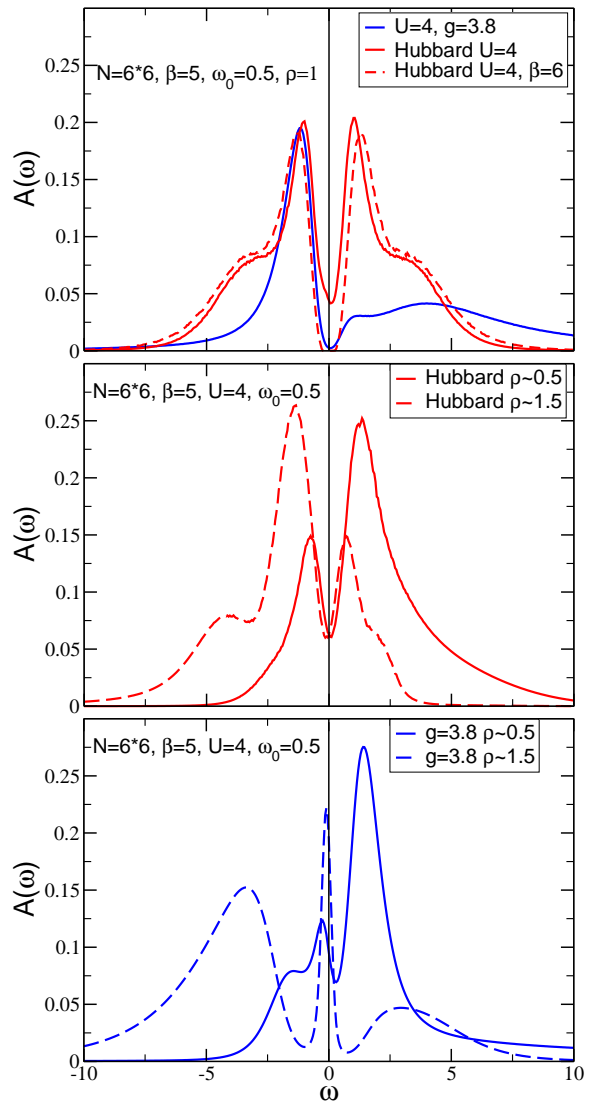

\centerline{\epsfig{figure=fig7.eps,width=7.5cm,clip}}
\centerline{\epsfig{figure=fig8.eps,width=7.5cm,clip}}
\hskip+0.11cm{\epsfig{figure=fig9.eps,width=7.62cm,clip}}
\caption{(Color online) Top: Comparison of the spectral function
$A(\omega)$ for the static and dynamic Hubbard models at $\beta=5$ and
half-filling.  We can see clearly that the Mott gap is more robust in
the dynamic case.  Middle and bottom: The behavior of $A(\omega)$ away
from half-filling\cite{note3}.  In all cases the spectral function is
finite at the Fermi energy, indicating metallic behavior.  However,
for the dynamic model at $\rho=1.5$ there is a sharp resonance at
$\omega=0$ whereas in the other cases the spectral function is
suppressed there.}
\label{AomegaRho1.0+0.5}
\end{figure}

The spectral function $A(\omega)$, which we obtain with an analytic
continuation of $G(\tau)$ using the maximum entropy
method\cite{maxent}, shows supporting evidence for the enhancement of
the Mott gap at half-filling, Fig.~\ref{AomegaRho1.0+0.5}(top).  Above
half-filling $A(\omega)$ exhibits a sharp resonance at $\omega=0$,
Fig.~\ref{AomegaRho1.0+0.5}(bottom).  The comparison of $A(\omega)$
for $\rho=0.5$ and $\rho=1.5$ further emphasizes the lack of
particle-hole symmetry, Fig.~\ref{AomegaRho1.0+0.5}(bottom).

\subsection{Pairing Susceptibilities}

We turn now to a discussion of superconductivity in the dynamic model.
In the static Hubbard model, it has been shown that the $s$-wave
pairing vertex is repulsive (positive).  The $d$-wave
vertex is negative, but only relatively weakly so at the temperatures
accessible to the simulations\cite{white89,white89b}.  Near
half-filling, the extended $s$-wave vertex is also attractive, but
markedly less so than $d$-wave, suggesting  that $d$-wave
symmetry is the most likely instability.  However, the same sign
problem which precludes a definitive statement about superconductivity
in the static model also limits what we can conclude here for the
dynamic model.  Nevertheless, there is an interesting qualitative
difference between the two models which can be clearly discerned.

\begin{figure}
\centerline{\epsfig{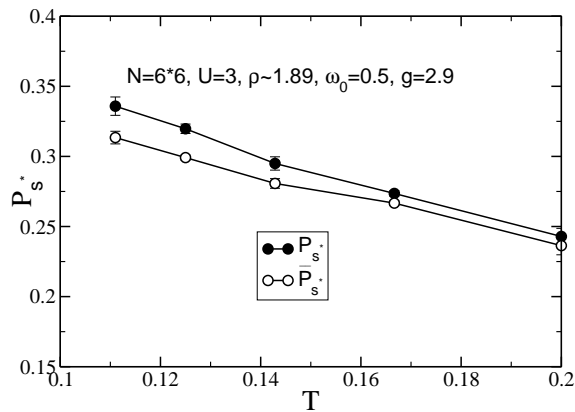}}
\caption{(Color online) The extended $s$-wave pair susceptibilities
$P_s^*$ and $\overline{P}_{s^*}$ as a function of temperature for
$U=3, \omega_0=0.5$, and $g=2.9$.  Here, unlike the static model,
$P_{s^*}$ exceeds $\overline{P}_{s^*}$ when the temperature is
lowered.  However, we cannot say if $P_{s^*}$ might diverge at low
temperature because of the sign problem.}
\label{PsxU3}
\end{figure}

\begin{figure}
\centerline{\epsfig{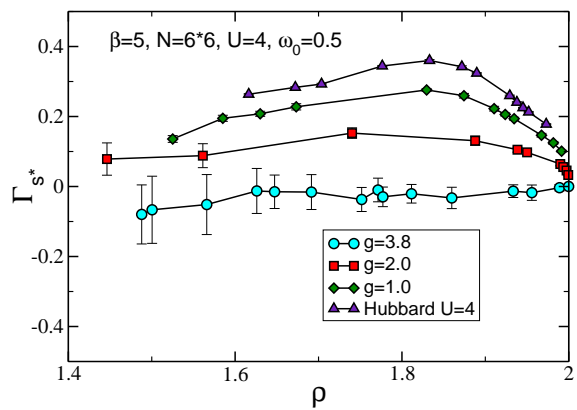}}
\caption{(Color online) $\Gamma_{s*}$ as a function of density $\rho$
for $\beta=5$ near full-filling. The $s^*$ channel becomes attractive
when $g$ increases.  $\Gamma_{s*} \rightarrow -1$ would signal a
superconducting instability.}
\label{Gammasx}
\end{figure}

Specifically, the extended $s$-wave vertex is attractive in the
dynamic model in the regime of $g$ where $U_{\rm eff}$ is negative,
while it is repulsive in the static model at these high fillings.
In Fig.~\ref{PsxU3} we compare the temperature evolution of the
correlated and uncorrelated susceptibilities, $P_{s^*}$ and
$\overline{P}_{s^*}$, at $\rho =1.89$ and $U=3$ and see that the $P_{s*} >
\overline{P}_{s^*}$.  
The average sign takes the values 0.94, 0.92, 0.83, 0.73, and 0.63 
at $\beta=5, 6, 7, 8, 9$ respectively.
The resulting attractive (negative) vertex is
given in Fig.~\ref{Gammasx}.  For $g=0$, the static model, the vertex
is repulsive.  But it systematically decreases and goes negative as
the coupling to the dynamic field is strengthened.  In this plot the
inverse temperature is fixed at $\beta=5$ and the density is allowed
to vary.
The average sign takes the values 0.36, 0.54, 0.73, 0.89, and 0.96
at $\rho = 1.50, 1.60, 1.70, 1.80, 1.90$.

\begin{figure}
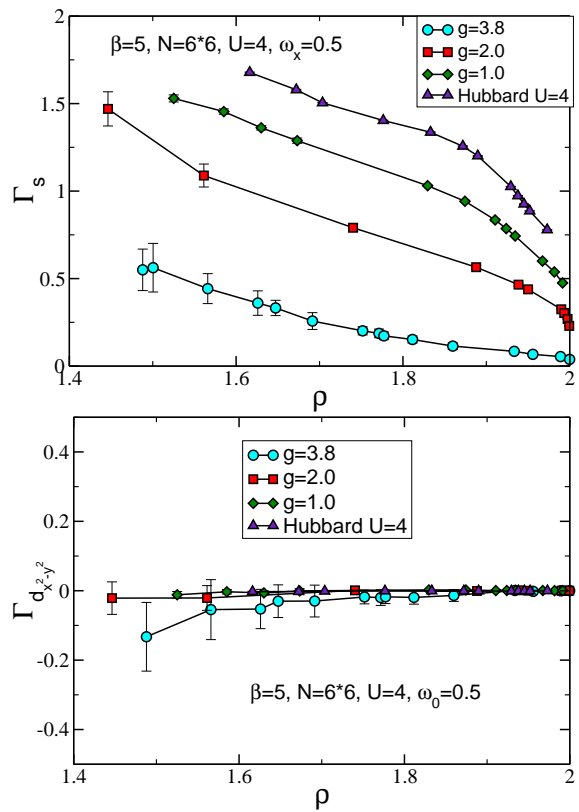

\centerline{\epsfig{figure=fig12.eps,width=7.5cm,clip}}
\centerline{\epsfig{figure=fig13.eps,width=7.5cm,clip}}
\caption{(Color online) Top: $\Gamma_{s}$ as a function of $\rho$ for
different values of $g$ at $\beta=5$.  Unlike $\Gamma_{s^*}$, the
$s$-wave channel remains repulsive.  Bottom: $\Gamma_{d_{x^2-y^2}}$ as
a function of $\rho$.  The $d_{x^2-y^2}$-wave channel is attractive,
but the effect is less pronounced than for $s^*$, especially near full
filling.}
\label{Gammasd}
\end{figure}

Figure~\ref{Gammasd} (top) shows that, in contrast to the behavior of
$\Gamma_{s^*}$, the s-wave vertex is strongly repulsive, although $g$
does weaken the repulsion somewhat as it increases.  Meanwhile, we see
in Fig.\ref{Gammasd} (bottom) that near full filling the $d$-wave
vertex is more weakly attractive than the $s^*$-wave vertex.  This
suggests that if the dynamic Hubbard model does have a superconducting
instability at small hole-doping that it would be of $s^*$ symmetry,
unlike the $d$-wave symmetry which is most attractive for the static
model\cite{note2}.

\begin{figure}
\centerline{\epsfig{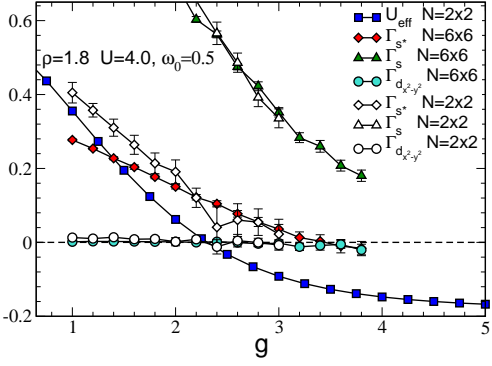}}
\caption{(Color online) Pairing vertices and binding energy as
functions of the dynamical coupling $g$.  $\Gamma_{s^*}$ becomes
negative at $g > 3.5$ while $U_{\rm eff} < 0 $ when $g > 2.3$.  As
long as $g<4$, i.e.~for the entire range of the horizontal axis, both
$U_{\rm min}$ and $U_{\rm max}$ are repulsive.  $\Gamma_{s^*}$ is
strongly repulsive in the static model $g=0$.  }
\label{UeffandGammas}
\end{figure}

It is informative to compare the onset of attraction in the
pairing vertex with the development of negative binding energy.
Fig.~\ref{UeffandGammas} shows $\Gamma_s, \Gamma_{s^*}$ and
$\Gamma_{d_{x^2-y^2}}$ versus $g$ for $U=4$ and $\omega_0=0.5$.  The
filling $\rho=1.8$.  On 2x2 lattices, for which the ED calculation of
$U_{\rm eff}$ is feasible, $\Gamma_{s^*}$ becomes negative at somewhat
larger values of $g$ than where $U_{\rm eff}$ becomes negative.  The
figure also shows that $\Gamma$ is relatively insensitive to lattice
size: the 2x2 and 6x6 lattices give results which are quantitatively
rather similar for most values of $g$.  Note also that $\Gamma_{s^*}$
is strongly repulsive in the static model $g=0$.

\begin{figure}
\centerline{\epsfig{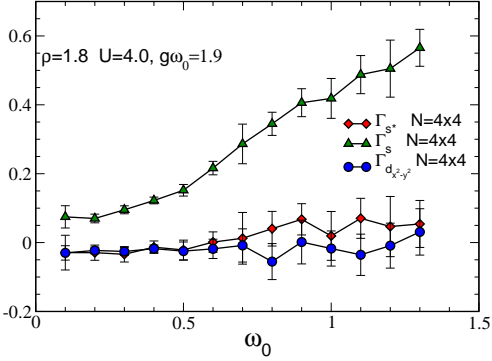}}
\caption{(Color online) Pairing vertices as
functions of frequency $w_0$ at fixed $g w_0 = 1.9$, $\rho=1.8$,
$\beta = 5$, 
and $U=4$.  The attractive d-wave vertex shows only a weak dependence on
$w_0$.
}
\label{Gammavsw0}
\end{figure}

A significantly larger enhancement of superconductivity was reported
\cite{marsiglio90} in a Hubbard Hamiltonian in which the hopping of one
spin species is modulated by the density of the other.  This model was
argued to be connected to the dynamic Hubbard Hamiltonian in the limit
of large $w_0$.  We conclude this section by exploring the $w_0$
dependence of the pairing vertex, to see if larger $w_0$ might show a
greater tendency for superconductivity.  In Fig.~\ref{Gammavsw0} we show
the vertices as a function of $w_0$.  We have fixed the product $g w_0 =
1.9$ and $U=4$ so we can stay near the values of $U_{\rm min}$ where the
binding energy is maximized. The attraction does not seem to
increase markedly with $w_0$.

\subsection{Energy}

\begin{figure}
\centerline{\epsfig{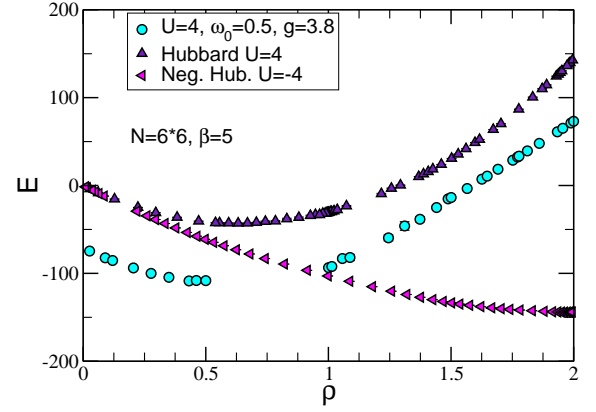}}
\caption{(Color online) Total energy as a function of $\rho$ at
$\beta=5$ for the static attractive and repulsive Hubbard models, and
for the dynamic model.  The static models show clear positive
curvature, indicative of thermodynamic stability.  In the dynamic
model $E(\rho)$ is nearly linear.}
\label{Etotfull}
\end{figure}

\begin{figure}
\centerline{\epsfig{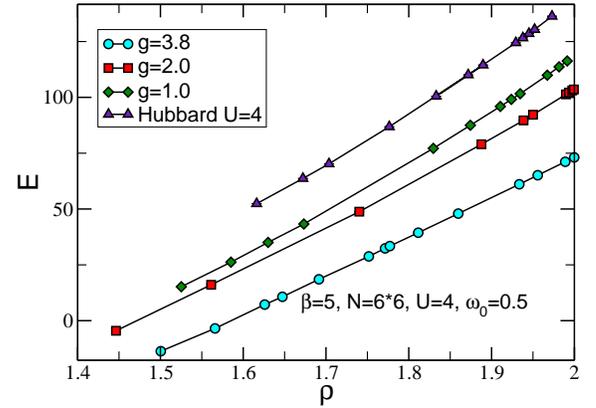}}
\caption{(Color online) Total energy as a function of $\rho$ at
$\beta=5$.  For $g=0$, the static model, the curvature is positive.
As $g$ gets larger, the energy becomes linear in the density.}
\label{Etot}
\end{figure}

The total energy (Fig.\ref{Etotfull}) also shows a markedly different
dependence on the density, $\rho$, in the dynamic Hubbard Hamiltonian.
Whereas the static positive and negative $U$ Hubbard Hamiltonians have
$d^2 E / d\rho^2 > 0$, the positive curvature of the dynamic model
that is evident below half-filling becomes very small for $\rho>1$ as
$g$ increases and eventually the curvature nearly vanishes.  Figure~\ref{Etot} shows this
linear behavior developing with $g$. 

The temperatures at which we performed our simulations
are low enough that the total internal energy, $E$, is nearly equal to the
free energy, $F$.  As it is well known, negative curvature in the
free energy as a function of the density, in the canonical ensemble,
leads to negative compressibility and is thus a signal for phase
separation and a first order phase transition\cite{phasesep}.
Thermodynamic stability requires positive curvature for the free
energy versus density. While our simulations are performed in the
grand canonical ensemble, where such negative curvatures are not
observed, we do see (Fig.~\ref{Etot}) a progression from positive to
zero curvature as $g\to 4$. At the same time, and recalling that
$\mu=\partial (F/V)/\partial \rho$, we see in Fig.~\ref{Murho} that as
$g\to 4$, the $\rho$ versus $\mu$ curves get steeper signalling higher
compressibility $\kappa=\partial \rho/\partial \mu$. Noting that
$U_{\rm min}$ vanishes for $g=4$ and becomes negative when $g>4$, we
interpret these observations as a possible phase separation setting in
at $g=4$ whereby the system develops hole-rich and hole-deficient
regions.

\section{Conclusions}

In this paper we have performed determinant Quantum Monte Carlo
simulations of a two dimensional Hubbard Hamiltonian in which the
on-site repulsion is coupled to a fluctuating bosonic field.  Our
studies complement earlier work using the Lang-Firsov transformation
and exact diagonalization and QMC in one dimension.  We note a number
of interesting features of the model.  First, the Mott gap at
half-filling is stabilized.  Second, antiferromagnetic correlations
are enhanced above half-filling.  The extended $s$-wave pairing
vertex, which is repulsive in the ordinary static Hubbard Hamiltonian,
is made attractive in the dynamic model.  The value of $g$ for which
this attraction manifests is roughly consistent with the value at
which the binding energy $U_{\rm eff}$ goes negative on 2x2 clusters.
The sign problem prevents simulations at low temperatures to see if an
actual pairing instability occurs.  We have also
observed that as $g\to 4$, {\it i.e.} as $U_{\rm min}\to 0$, $E(\rho)$
becomes linear in $\rho$ signalling possible phase separation into
regions of hole-deficient and hole-rich regions when $U_{\rm min}$
becomes negative for $g>4$.  Finally, we note that we have also found,
within the Hartree-Fock framework, that charge inhomogeneities
(stripes) are supported by this dynamic Hubbard model \cite{bouadimunp}.

KB, FH, and GGB acknowledge financial support from a grant from the
CNRS (France) PICS 18796, RTS from NSF ITR 0313390, and ME from the
Research Corporation.  We acknowledge very useful help from M. Schram
and R.Waters.


\begin{thebibliography}{10}
\bibitem{note3} In the static Hubbard case, Fig.~\ref{AomegaRho1.0+0.5}(middle),
the densities are $\rho=0.497$ and $\rho=1.502$
which is one cause of the slight asymmetry.
\bibitem{note2} We comment that the attraction exhibited in the
$d$-wave vertex of the static model is most manifest at half-filling,
where the cuprates are antiferromagnetic insulators. This is possibly,
however, a consequence of small lattice size since then the pair
creation operator $\Delta^\dagger$ which adds two particles to the
system, significantly shifts the density.
\end{thebibliography}
\end{document}